\documentclass[aps,prl,reprint,superscriptaddress,showpacs]{revtex4-1}

\usepackage{graphicx}
\usepackage{color}
\usepackage{booktabs}
\usepackage{bm}
\usepackage{longtable}
\usepackage{amsmath}
\usepackage{dcolumn}
\usepackage{placeins}
\usepackage{expdlist}
\usepackage{physics}

\begin{document}

\newcommand{\avg}[1]{\langle #1 \rangle}
\newcommand{\up}{\uparrow}
\newcommand{\down}{\downarrow}
\newcommand{\eff}[1]{#1_{\rm eff}}
\renewcommand{\Re}{\mbox{Re}}
\renewcommand{\Im}{\mbox{Im}}


\title{Optical control and coherent coupling of spin diffusive modes in thermal gases}

\author{P. Bevington$^{\dagger}$}
\affiliation{National Physical Laboratory, Hampton Road, Teddington, TW11 0LW, United Kingdom}
\author{J. Nicholson$^{\dagger}$}
\affiliation{School of Physics and Astronomy, University of Birmingham, Edgbaston, Birmingham B15 2TT, United Kingdom}
\author{J. D. Zipfel}
\author{W. Chalupczak}
\affiliation{National Physical Laboratory, Hampton Road, Teddington, TW11 0LW, United Kingdom}
\author{C. Mishra}
\author{V. Guarrera}
\email{v.guarrera@bham.ac.uk}
\affiliation{School of Physics and Astronomy, University of Birmingham, Edgbaston, Birmingham B15 2TT, United Kingdom}

\def\thefootnote{$\dagger$}\footnotetext{These authors contributed equally to this work.}


\begin{abstract}
Collective spins in thermal gases are at the core of a multitude of science and technology applications. In most of them, the random thermal motion of the particles is considered detrimental as it is responsible for decoherence and noise. In conditions of diffusive propagation, thermal atoms can potentially occupy various stable spatial modes in a glass cell. Extended or localized, diffusive modes have different magnetic properties, depending on the boundary conditions of the atomic cell, and can react differently to external perturbations. Here we demonstrate that few of these modes can be selectively excited, manipulated, and interrogated in atomic thermal vapours using laser light. In particular, we individuate the conditions for the generation of modes that are exceptionally resilient to undesirable effects introduced by optical pumping, such as light shifts and power-broadening, which are often the dominant sources of systematic errors in atomic magnetometers and co-magnetometers. Moreover, we show that the presence of spatial inhomogeneity in the pump, on top of the random diffusive atomic motion, introduces a coupling that leads to a coherent exchange of excitation between the two longest-lived modes. Our results indicate that systematic engineering of the multi-mode nature of diffusive gases has great potential for improving the performance of quantum technology applications based on alkali-metal thermal gases, and promote these simple experimental systems as versatile tools for quantum information applications.

\end{abstract}

\maketitle

\section{Introduction}
Due to their long-lived collective spin states, gaseous mixtures of alkali-metal and noble-gas atoms are widely applied in quantum optics and sensing. At room temperature and above, atoms in a low-pressure gas enclosed in a glass cell move along ballistic trajectories at velocities on the order of hundreds of meters per second. Alkali-metal spin ensembles in these regimes rapidly reach the cell walls, where they get depolarized. To reduce this effect and increase the alkali-metal spins' polarization lifetime, it is common practice to fill the cell with $10^{-1}$ to $10^4$ torr of a buffer gas, typically an inert or diatomic gas with low polarizability. Velocity-changing collisions with the buffer gas dramatically modify the thermal motion of the spins, which becomes diffusive. In these conditions, atoms can occupy different, stable spatial modes. Few experiments have reported diffusion effects in the context of light storage and slowing-down \cite{Shuker08, Firstenberg10}, and atomic (non-spin) coherence \cite{Xiao06}. Even though signatures of diffusion have been detected in optically pumped atomic magnetometers \cite{Skalla97, Knappe10}, the impact of higher order modes on the collective spin dynamics has not been systematically analysed. It has been recently realised, however, that accessing and controlling diffusive spatial modes can provide great advantages for systems based on atomic thermal vapours, e.g. for reducing noise and instability in ultra-sensitive atomic sensors \cite{Xiao2023}, for the realization and study of complex non-Hermitian systems \cite{Li2019}, for the development of efficient spins systems with application in quantum information \cite{Shaham22}, and the generation of non-local classical and non-classical correlations \cite{Shaham20,Katz22}.  

In the present paper, we access and characterize the underlying multi-mode dynamics of an atomic system in a range of parameters which is common for atomic magnetometry and co-magnetometry applications. By using spatially resolved non-destructive spin imaging \cite{Dong19,Xia06}, we can identify few stable spatial modes with different magnetic properties that correspond to low-order diffusive modes of the alkali-metal ensemble. Our measurements show that, since pump and probe beams overlap differently with the different spatial modes, the collective spin dynamics of the latter develops features that crucially depend on the parameters of the light beams, such as intensity and beam geometry. In analogy with the Ramsey-narrowing effect seen in EIT systems \cite{Xiao06}, we can create long-lived collective spin modes for which the sensitivity to light shifts and power broadening is reduced by up to one order of magnitude with respect to standard pumping schemes. We also show that spatial modes with different magnetic properties can coherently exchange collective excitation. Interestingly, in the presence of a noble-gas buffer, the inhomogeneous optical pumping can turn the coupling between different alkali-metal spatial modes from mainly incoherent to coherent. This is remarkable, especially when considering that the mechanism underlying the coupling is based on a random diffusive dynamics. Our results demonstrate that the multi-mode approach to the spin dynamics is a powerful tool to enhance the performance of spin vapour devices, such as those used for sensing applications. The realization of optically accessible, localized spin modes opens also exciting perspectives in the wider scenario of quantum information and imaging.

\section{Results}

\subsection{Stationary spatial-modes in hot atomic vapours}
\begin{figure}
\centering
\includegraphics[width=0.35\textwidth]{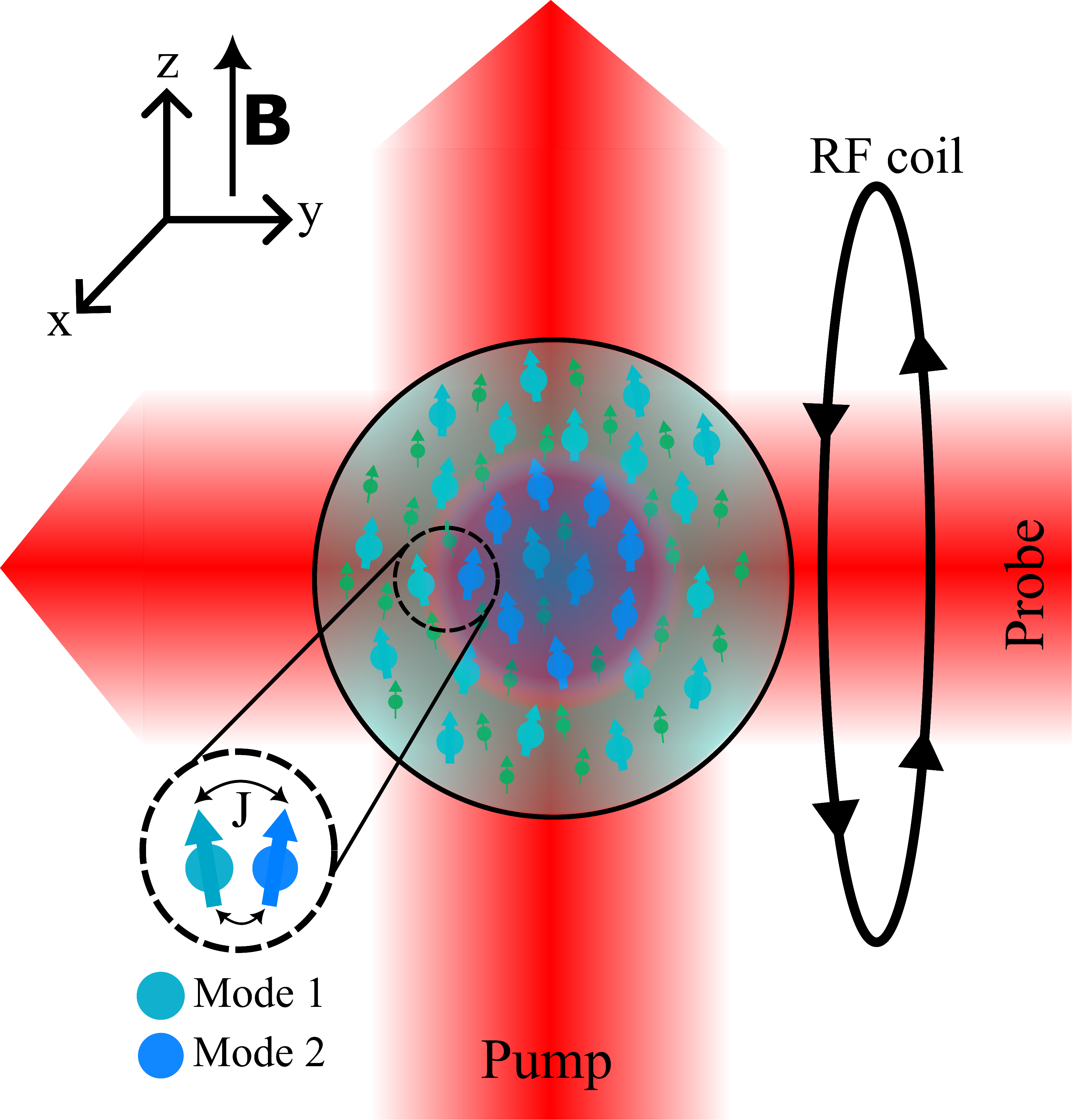}
\includegraphics[width=0.4\textwidth]{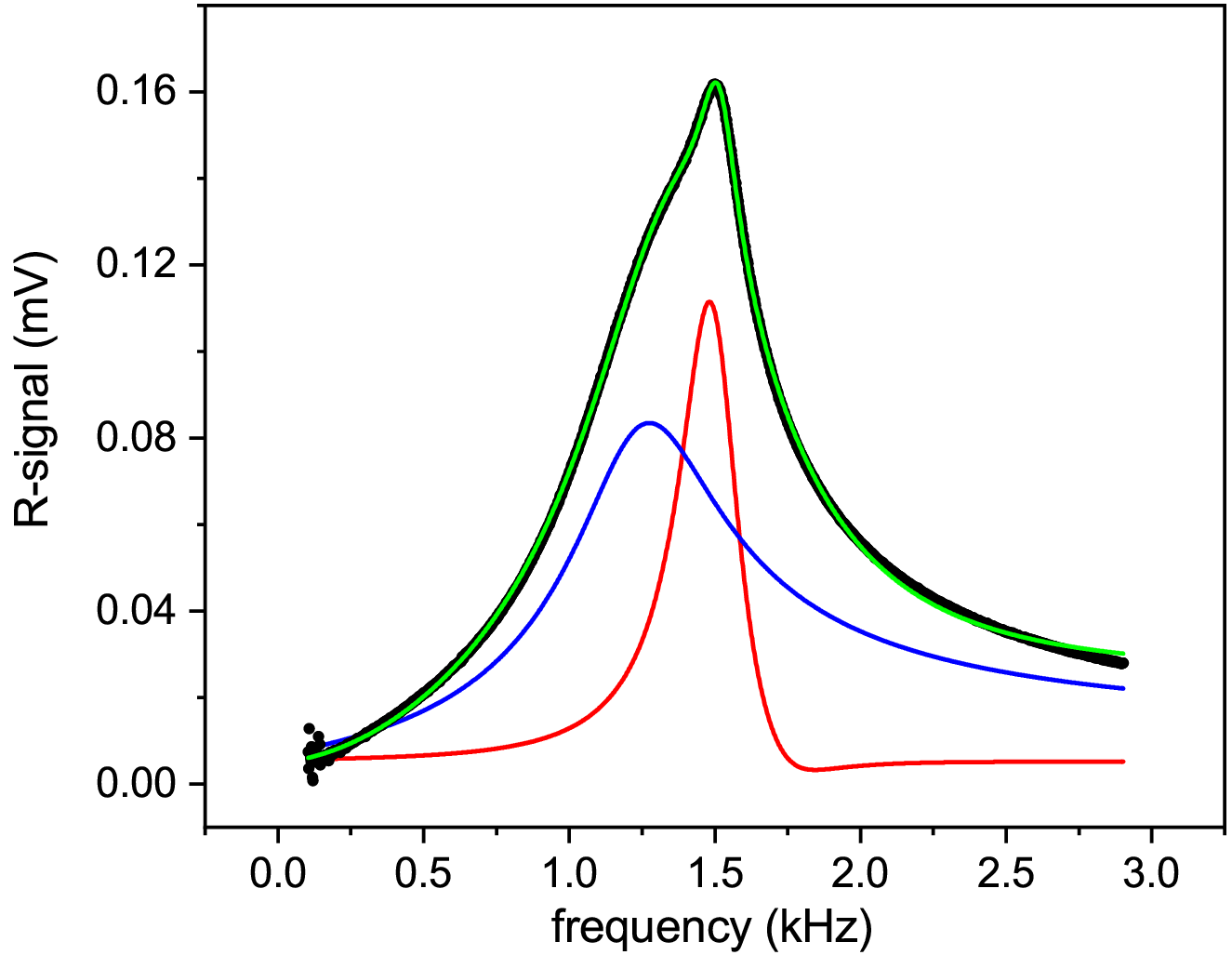}
\caption{Up: schematics of the experiment for excitation and probe of the alkali-metal/noble gas mixture. Down: measured spin-rotation signal, obtained from a lock-in amplifier referenced at the RF frequency, showing a composite line shape. The fit (solid green line) is done by using two Lorentzian functions (solid red and blue lines). Fitted parameters are: amplitude, linewidth, resonance frequency, and phase of the two Lorenztian line shapes, with an off-resonant background. }\label{fig:fig1}
\end{figure}

We analyse the Faraday-type polarization rotation signal generated by the transverse spin excitation of an alkali-metal gas in a spherical cell with 500 torr of neon buffer gas (Twinleaf). The alkali spins are polarized along the axial magnetic field by an optical pumping beam (sigma-polarized and resonant with the $F=3 \longrightarrow F'=2$ transition for cesium), and transverse excitation is generated by a weak RF resonant field, see Fig.\ref{fig:fig1}. 
The neon gas is partially polarized by spin-exchange collisions with the polarized alkali atoms (a process called spin-exchange optical pumping or SEOP), and has an axial coherence time longer than 30 minutes, thus its individual relaxation during the measurements is hereafter neglected. The cell also contains 20 torr of nitrogen gas for quenching of the alkali excited states. The dynamics of the alkali transverse excitation is monitored for roughly $1$ second after switching on the pump beam and the RF modulation. The signal is obtained by analyzing the polarization of a probe beam (linearly polarized and $> 15$ GHz detuned from all $F=4$ transitions), and is processed by a lock-in amplifier. We can perform both integrated and spatially-resolved measurements. The latter are taken by shifting a mm-sized slit over the probe beam profile, which allows a mm-resolution to be obtained in the central region of the cell.
 
\begin{figure}
\centering
\includegraphics[width=0.45\textwidth]{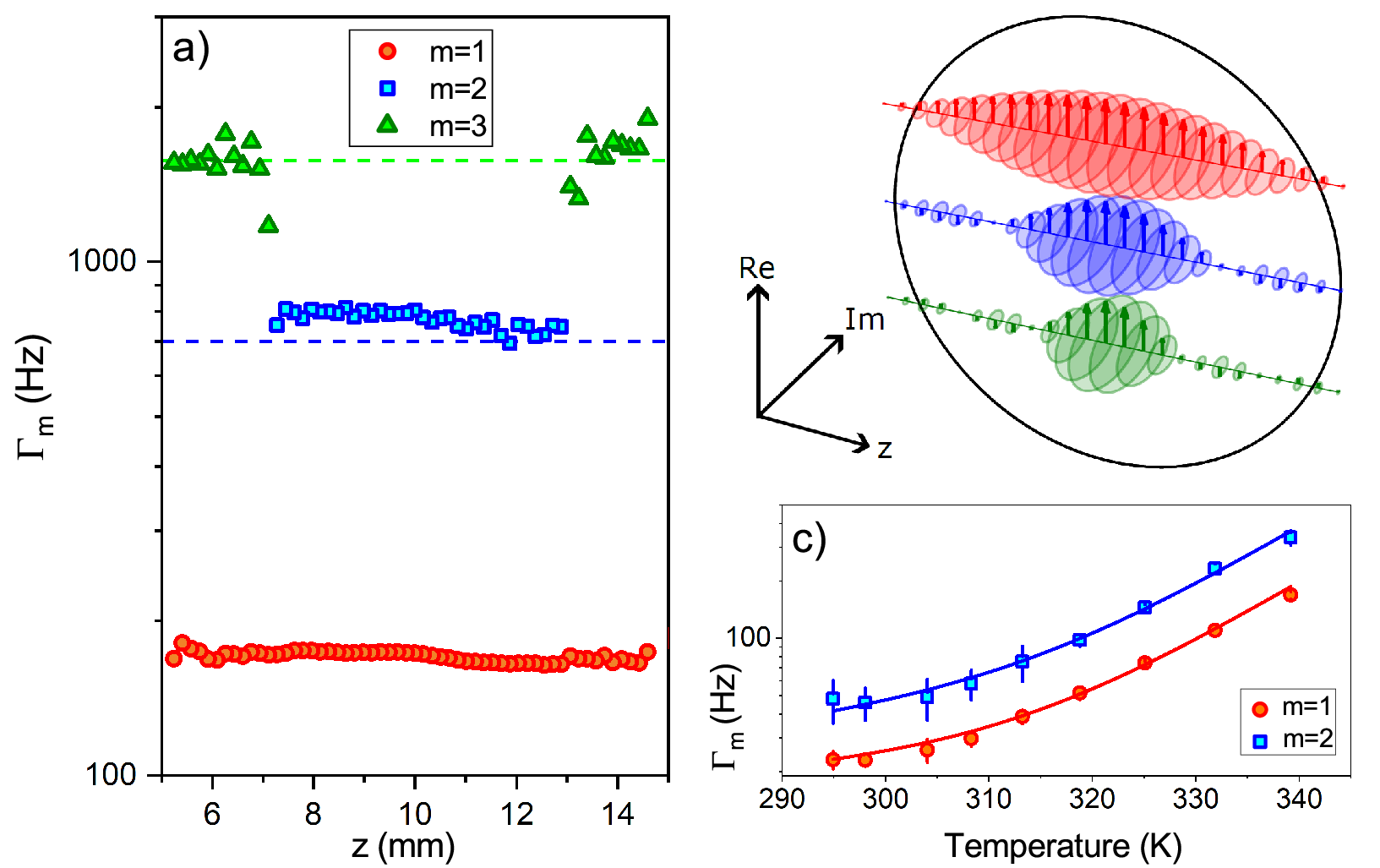}
\includegraphics[width=0.45\textwidth]{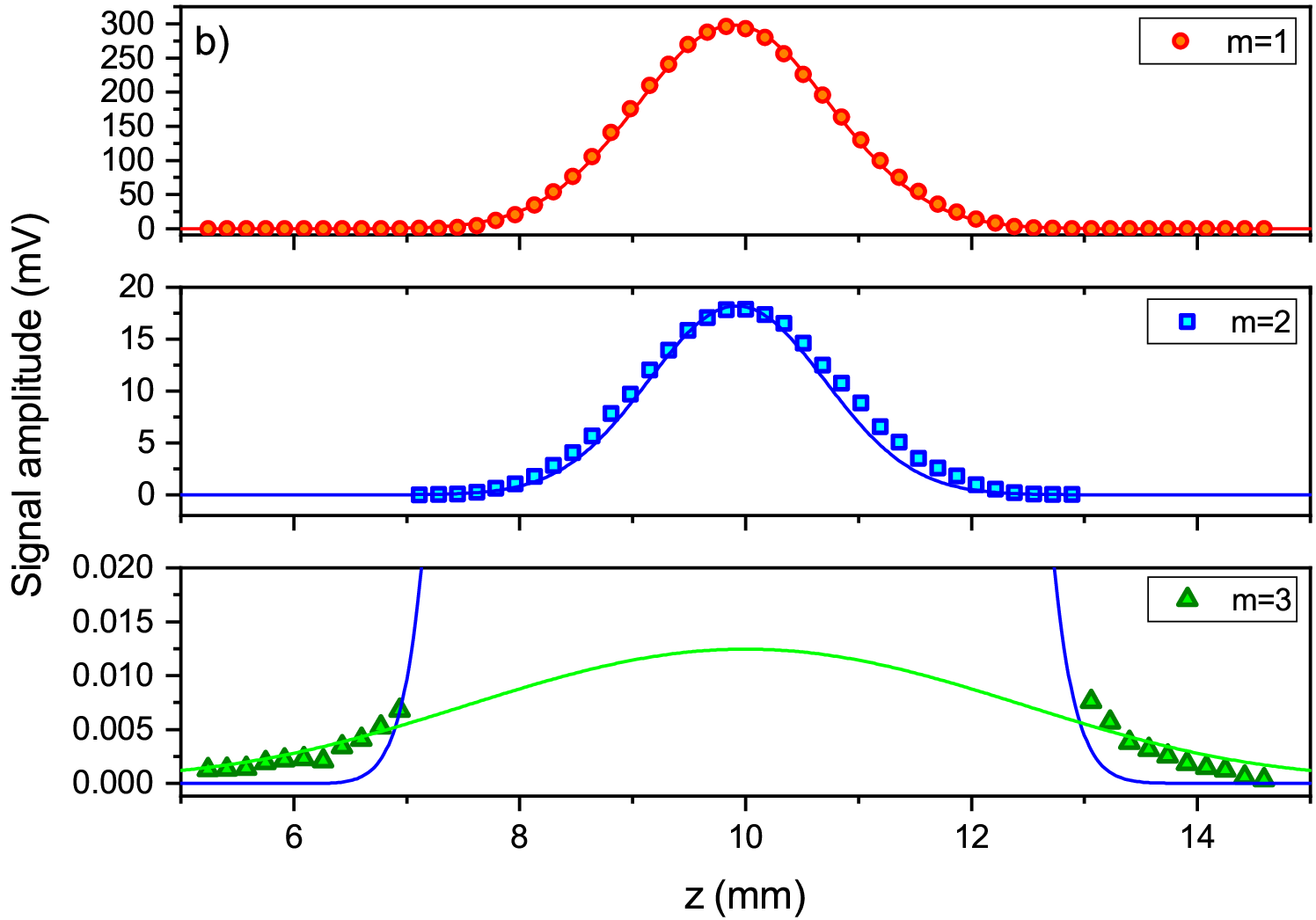}
\caption{a) Linewidths of the two main Lorentzian components of the spatially resolved spin-rotation signal along the $z$-axis. The measurements show the presence of three distinct stable spatial modes with characteristic magnetic features. The data, obtained in a regime of dirty-cell (see text), provide $\Gamma_2/\Gamma_1=4.4(2)$ and $\Gamma_3/\Gamma_1=9.6(6)$. Measurements are then compared to the low-order diffusive modes of the atomic cell with largest coupling to the pump Gaussian profile (in the cartoon inset $s_{000},s_{100}, s_{200}$ along the $z$-axis). Specifically: b) signal amplitude for each fitted mode as a function of the probe position along the $z$-axis. Solid lines are fits obtained by convolution of the diffusive modes with the profile of our probe beam; c) temperature dependence of the $m=1,2$ modes signal linewidth measured for low pump powers ($\lesssim 30$ $\mu$W). Solid lines are fits based on an estimated offset with the temperature dependent model discussed in the text. The ratio between the modes' linewidth is compatible with predictions for the lowest order diffusive modes, also in a regime where diffusion through the cell is not the main effect limiting the coherence lifetime, but it has a temperature dependence associated with a change in the cell boundary conditions. All measurements were taken with a low pump power of $0.3$ mW, and each point is an average over at least 20 different acquisitions.  }\label{fig:fig2}
\end{figure}

A careful inspection of the measured magnetic resonance line shape shows that it does not correspond to the single Lorentzian profile, which is typically associated with the lowest-order diffusion mode \cite{Knappe10,Xiao06}, see Fig.\ref{fig:fig1}. We interpret the observed additional features as different stationary spatial modes of the atomic gas, which we identify as higher-order diffusion modes of the alkali-metal atoms ensemble. The diffusive dynamics of the spin excitation is written in terms of the transverse magnetization $S_+=S_x+iS_y \equiv S$, and it is described by the equation:
\begin{equation}
\frac{\partial S}{\partial t} = D \nabla^2 S + \Gamma S
\label{eq1}
\end{equation}
where $D$ is the diffusion coefficient, and $\Gamma=D k^2$ the relaxation rate due to collisions with the cell walls and diffusion effects. 
Stationary spatial modes, $s_{nlp}(\textbf{r})$, can be obtained by solving the Helmholtz equation above with the Robin boundary condition \cite{Wu88,Masnau67,Shaham20}. For a spherical cell geometry, the boundary condition is thus set by the equation:
\begin{equation}
-\frac{j_{l}(k_{nl}R)}{j^{'}_{l}(k_{nl}R)}=\frac{2}{3} \frac{(1+e^{-1/N})}{(1-e^{-1/N})}\lambda k_{nl}
\end{equation}
and $s_{nlp}(\textbf{r})=j_l(k_{nl}r)Y_{lp}(\theta,\phi)$, with $j_l(x)$ the spherical Bessel functions of the first kind, $R$ the cell radius, $\lambda$ the mean-free path, $N\lesssim 1$ in the absence of spin-preserving cell coating, and $Y_{lp}(\theta,\phi)$ the spherical harmonics. The evolution of $S$ can be expanded using the above modes with initial amplitude $c_{nl}$, calculated by convolution with the Gaussian spatial profile of the pump beam and assuming homogeneous pumping along the pump propagation direction, Fig.\ref{fig:fig2}(b). For simplicity of notation we reduce the index $(n,l,p)=m$, and we obtain:
\begin{equation}
S=\sum_m c_m s_m(\textbf{r}) e^{-(\Gamma_m+i \omega_m )t}
\end{equation} 
where $\omega_m$ is the precession frequency in the presence of a magnetic field, $\Gamma_m=D k_m^2+\Gamma$, and the decoherence rate $\Gamma= \frac{1}{\epsilon} \left( R^{SD}_{Cs-Ne}+R^{SD}_{Cs-N2}+R^{SD}_{Cs-Cs}+R^{SE}_{Cs-Ne}+R_P \right)$ $+\frac{1}{q_{SE}} R^{SE}_{Cs-Cs}+R_{grad}$ takes into account the spin destruction relaxations, spin exchange relaxations (alkali-alkali, alkali-noble gas), power broadening from the pump, and magnetic gradients \cite{ghosh09}. The slowing down factor $\epsilon$ depends on sample polarization, and $q_{SE}$ on the relative strength of the magnetic field $B$ and spin-exchange rate $R^{SE}_{Cs-Cs}$ \cite{Happer77}. 

The Faraday-type measurements are fitted with a sum of $\Tilde{n}$ Lorentzian functions, which leads to a very good agreement with $\Tilde{n}=2$, as shown in Fig.\ref{fig:fig1}. This allows the main modes' parameters to be extracted, such as the signal amplitude, decay rate, magnetic resonance frequency, and phase. The spatially-resolved measurements, see Fig.\ref{fig:fig2}(a), show the existence of three main modes with different $\Gamma_m$, and allow their spatial distribution to be retrieved along the z-axis (and integrated along the other two axis). A comparison with the theoretically derived diffusive modes provides a good quantitative agreement, once a Gaussian-shaped envelope for the probe beam is taken into account, Fig.\ref{fig:fig2}(b). We also measured the modes' linewidth as a function of the sample temperature. We note that all these measurements have been collected in the limit of small pump intensities, to neglect the effects of power broadening, and weak RF signal. We observe that the ratio between the linewidths of the first two modes does not appreciably change with temperature. For our system, this is possible if we include a dependence of the effective radius of the cell on the alkali-metal vapor pressure, i.e. if we allow the atoms deposited at the cell walls to have an impact on the coherence lifetime, e.g. via long range electrostatic interactions. In support of this statement, we have seen that the resonance linewidth irreversibly broadens with increasing temperature of the cell, which means that, once broadened, it does not regain its initial value even after the heating system is switched off. This excess broadening disappears only after the cell is removed from the apparatus and the glass is cleaned from the residual alkali-metal adsorbed at its surface. Fitting with a model which takes account of this effect is shown in Fig.\ref{fig:fig2}(c). Using the collisional cross sections and diffusion coefficients available in literature for the mixture cesium-neon (\cite{Lou18} and references therein), we could retrieve from the fit a ratio $k_2/k_1=1.9(1)$. This is compatible with the theoretical prediction of $2$ and with separate measurements made in the regime where the dominant source of spin destruction is interaction with the cell walls, for which we obtain $k_2/k_1=2.1(1)$ (and $k_3/k_1=3.1(2)$), see Fig.\ref{fig:fig2}(a).          

\begin{figure}
\centering
\includegraphics[width=0.4\textwidth]{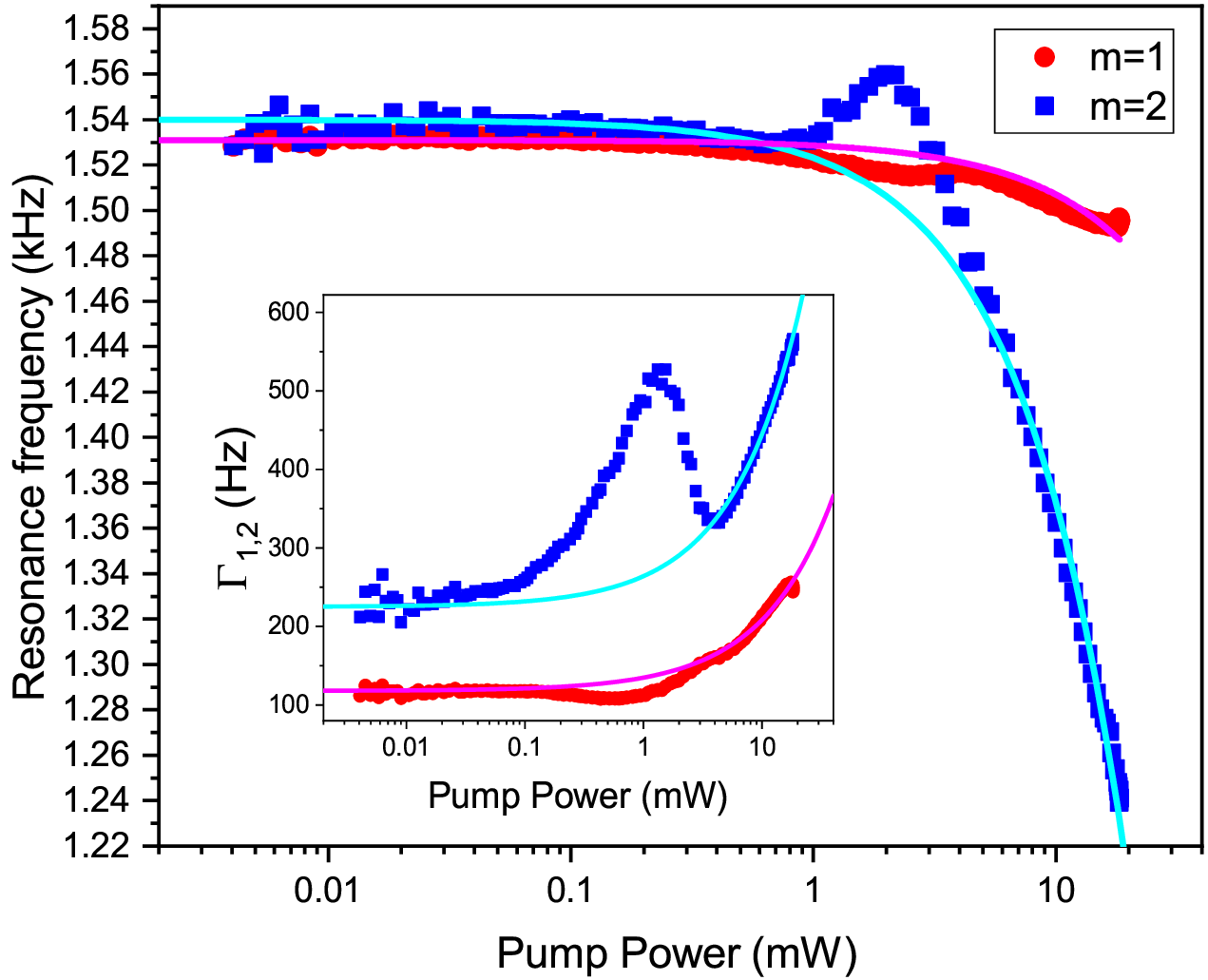}
\caption{Resonance frequency and linewidths (inset) of the modes $m=1,2$ as a function of the pump power. Solid lines are fits of the asymptotic behaviour, showing the light-induced shift and broadening of the two magnetic resonances. The lowest order mode is consistently less affected by the presence of the pump light, this depending on the actual size of the beam. The features visible around $2$ mW will be discussed later in the text. Each data point is the result of a fit over 100 different acquisitions.}\label{fig:fig3}
\end{figure}

As alkali-metal gases are routinely interfaced with laser beams for manipulation and readout, it is crucial to test the different features and behaviour of the spatial modes in the dependence of a controlled perturbation, like the one introduced by a pump beam. For high pump powers ($I\gtrsim I_{sat}$), we can extract independent information on the interaction between the pump and the spatial modes from the off-resonant power broadening (in the limit of low temperature $T\leq 30$ $^o$C) and light-induced Zeeman shift of the magnetic resonances \cite{Happer68} (see Fig.\ref{fig:fig3}). For the two modes $m={1,2}$, we obtain a similar functional dependence, but with an average mutual scaling factor of $I_2/I_1=7.8(4)$ and $I_2/I_1=8.4(3)$, respectively derived from the linewidth and the resonance position measurements. Notably, the main spatial mode $m=1$ is not only the longest-lived diffusive mode, but it is also the mode which is most resilient to the pump perturbation, without loss of signal amplitude. This effect can be explained as the atoms occupying the lowest-order diffusion mode move in and out of the pump beam, which has a Gaussian-like profile with a radius smaller than the radius of the cell \cite{Xiao06, Hunter22}. Both the pump and the probe beam can overlap with different spatial modes of diffusion. As a result, the properties of the detected signal, and its measured noise spectrum, depend on the spatial profile of the beams, the dimensions of the cell, and the diffusion characteristics of the atomic sample. We have verified, for example, that increasing the size of the pump beam changes which mode is dominant (for the same beam intensity), and can increase or suppress shielding from the light-induced perturbation (see supplemental material). This indicates that for the optimization of magnetometers and co-magnetometers, the size and shape of the pump beam should be carefully calibrated with respect to the extension and features of the diffusive modes supported by the cell geometry.   

\subsection{Coherent coupling of spatial modes}
Coherent and periodic exchange of spin excitation between an alkali-metal vapour and a noble gas has been recently demonstrated in the strong coupling regime \cite{Shaham22}. The different intra-species modes can be projected onto a basis of diffusive spatial modes, which contribute independently to the overall transverse spin dynamics \cite{Katz22}. Studies of the spatial modes' behaviour in the context of parity-time (PT) symmetry breaking have been very recently conducted in a system of nuclear spins, and improvement of atomic magnetometry has been demonstrated in a PT-broken phase \cite{Zhang2023}.       
\begin{figure}[h]
\centering
\includegraphics[width=0.45\textwidth]{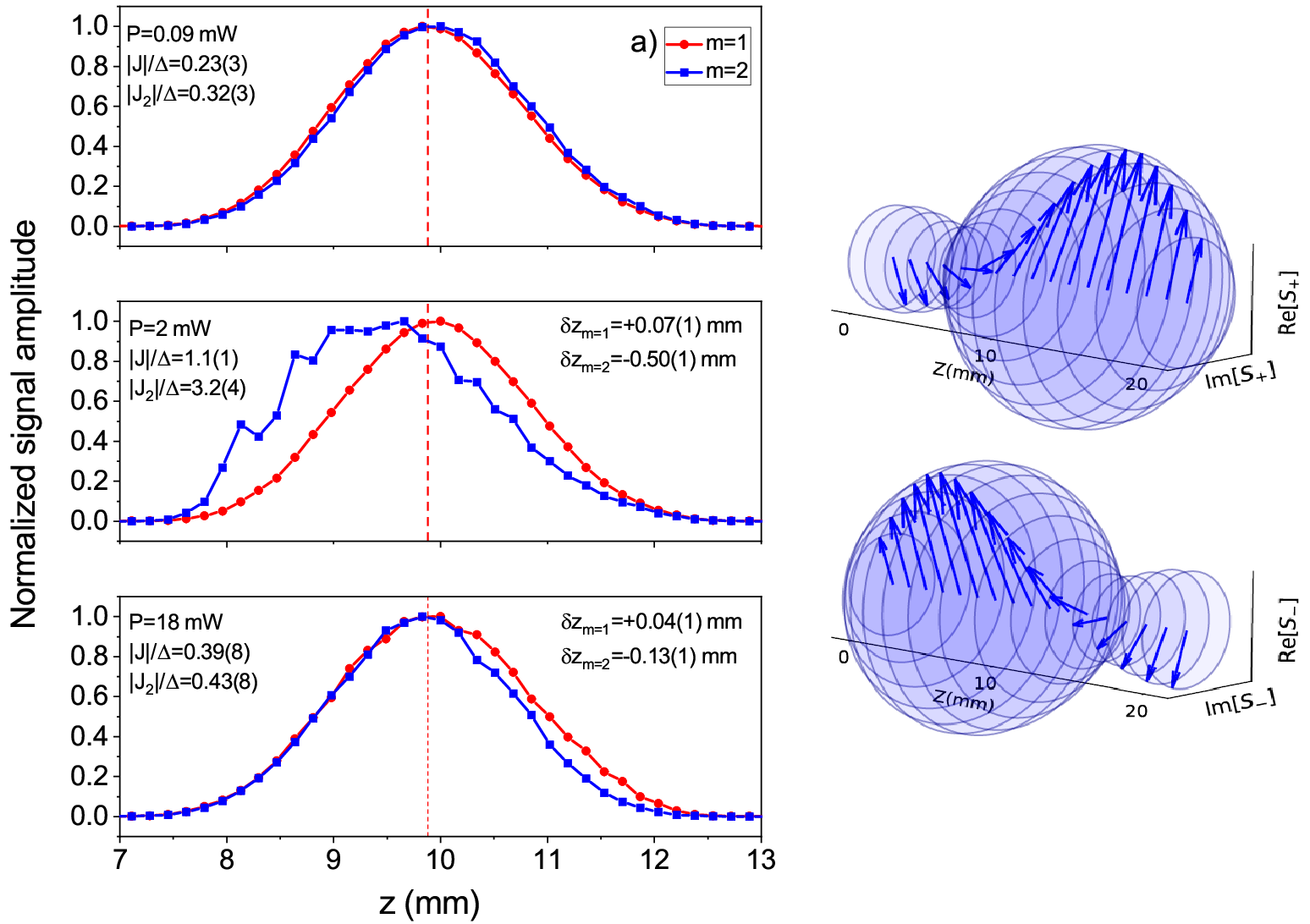}
\includegraphics[width=0.45\textwidth]{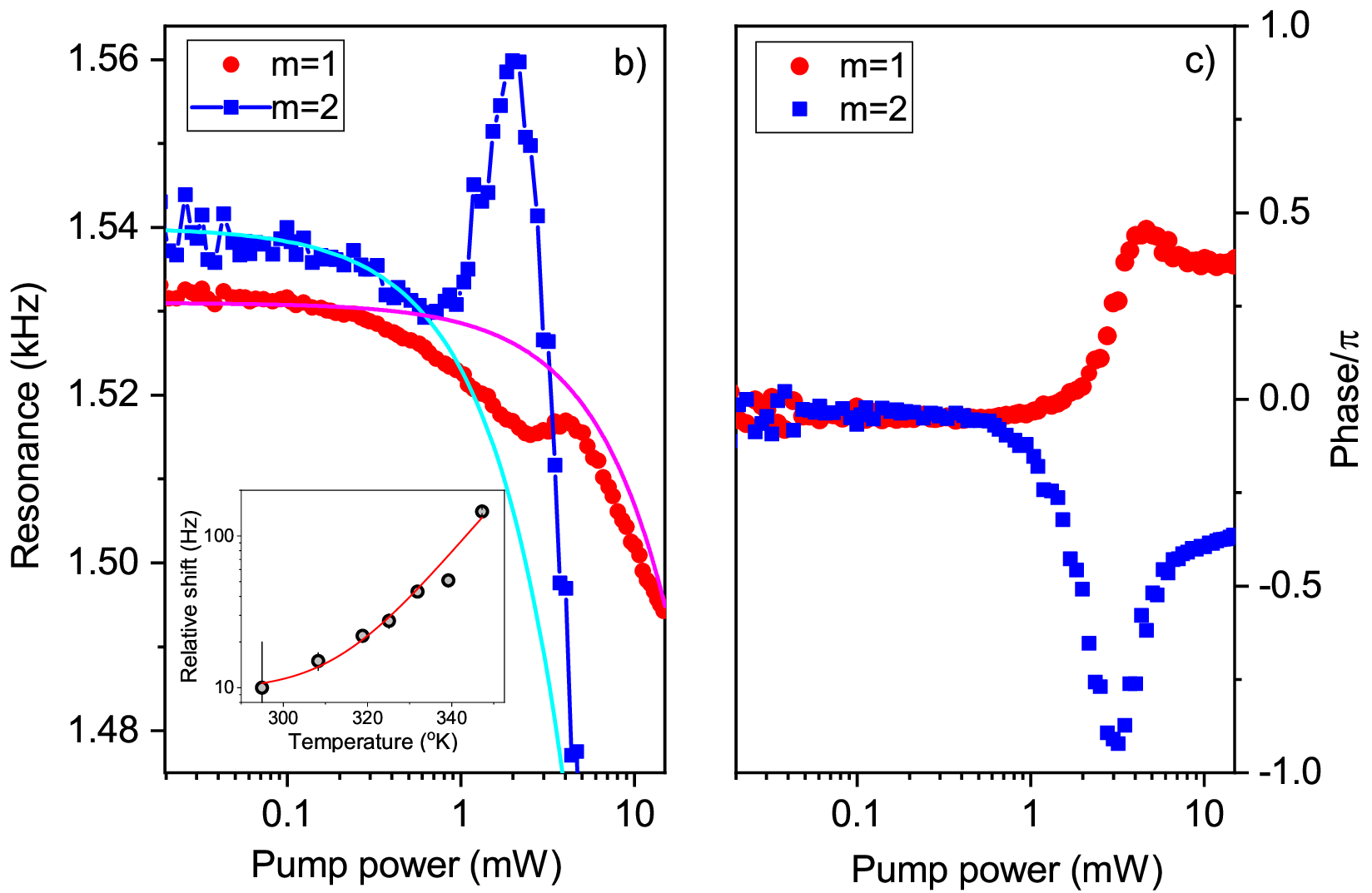}
\caption{a) Spatial imaging of the two main spin modes along the direction of the pump beam, recorded for different pump powers. Fitted displacement of the modes from their initial position, i.e. low pump power measurements, is shown in the picture, together with $\vert J_2 \vert,\vert J \vert$ obtained from the same spin imaging measurements (see also supplemental material). b) Zooms of the pump power dependence of the resonance frequency and phase show the two modes shifting in opposite directions, before the signal is mainly led by the light-induced shift (shown with solid lines). In the inset: peak resonance frequency separation between the modes $m=1,2$ (in the regime not dominated by the light-induced shift), as a function of temperature. Red solid line is a model of spin-exchange collisional shift. c) As the frequencies move apart, the phases of the two modes shift in opposite directions.}\label{fig:fig4}
\end{figure}

Different spatial modes of a single atomic species do not usually coherently couple, the solution of the uniform Bloch equation with diffusive dynamics providing a stable spatial orthogonal basis. However, in the presence of a spatial inhomogeneity, such as a magnetic field gradient or a light-shift gradient, a certain mixing of the different spatial modes might happen. In the simplifying assumptions of the spatial gradient adding to the Hamiltonian $H_0=D \nabla^{2}-i\gamma B -\Gamma=D \nabla^{2}-i\omega -\Gamma$ as a linear perturbation $H_{inh}=-i\gamma G z$ ($G$ generally being a complex number), and in the approximation of considering the two spatial modes with the longest lifetime: solutions of Eq.\ref{eq1} can be sought in the form $S(\textbf{r},t)=c_1(t) s_1(\textbf{r})+c_2(t) s_2(\textbf{r})$, where due to the pumping inhomogeneity, $s_1(\textbf{r})=s_{000}(r,\theta,\phi)$, and $s_2(\textbf{r}) = s_{011}(r,\theta,\phi)$. In this case, the coefficients $c_1, c_2$ evolve according to the equations (see also \cite{Zhang2023}):
\begin{equation}
    \begin{bmatrix}
        \dot{c}_1(t) \\
        \dot{c}_2(t) \\
    \end{bmatrix}
    = \left( E_0+ \Delta 
    \begin{bmatrix}
        1 & J/\Delta \\
        J/\Delta & -1 \\
    \end{bmatrix} \right)
 \begin{bmatrix}
        c_1(t) \\
        c_2(t) \\
    \end{bmatrix}
    \label{eq:eq2}
\end{equation}
with $E_0= \frac{(i \omega_1 - \Gamma_1 + i \omega_2 - \Gamma_2)}{2}$, $\Delta = \frac{(i \omega_1 - \Gamma_1  -i \omega_2 + \Gamma_2)}{2}$, and $J= -i \gamma G \int_{\Sigma} s_2^{*}(\textbf{r}) z s_1(\textbf{r}) d\textbf{r} $ is the coupling between the spatial modes due to the inhomogeneity. In the limit of $ \vert J/\Delta \vert >1$, the spatial modes of the transverse collective spins can become coherently coupled. In our case where $\omega_1 \sim \omega_2$ and $\Gamma_1 \sim \Gamma_2/2$, $\vert J \vert > \Delta \sim \Gamma_1/2$ and the effects of the coupling might become visible as an exchange of collective spin excitations between the two spatial modes \cite{Fang2021}. 

Exploiting the experimental access to different diffusive modes of the alkali-metal atoms, we analyze the transverse spin excitation induced by a weak RF field. For a fixed interrogation time (roughly $1$ sec), we vary the coupling term $J$ by changing the pump intensity. Indeed, at position $z$ in the cell, the features of the magnetic resonances (amplitude, frequency, linewidth) are mainly determined by the competition between optical pumping and spin-exchange processes. For example, depending on the dominant process, the precession frequency is mainly shifted by light, or by spin-exchange interaction with the noble gas magnetization $\textbf{M}$ (with opposite sign). Thus in an optically thick cloud, the competition between these different effects results in a spatial gradient of the frequency shift (and more in general, of all the magnetic properties).   
\begin{figure}
\centering
\includegraphics[width=0.45\textwidth]{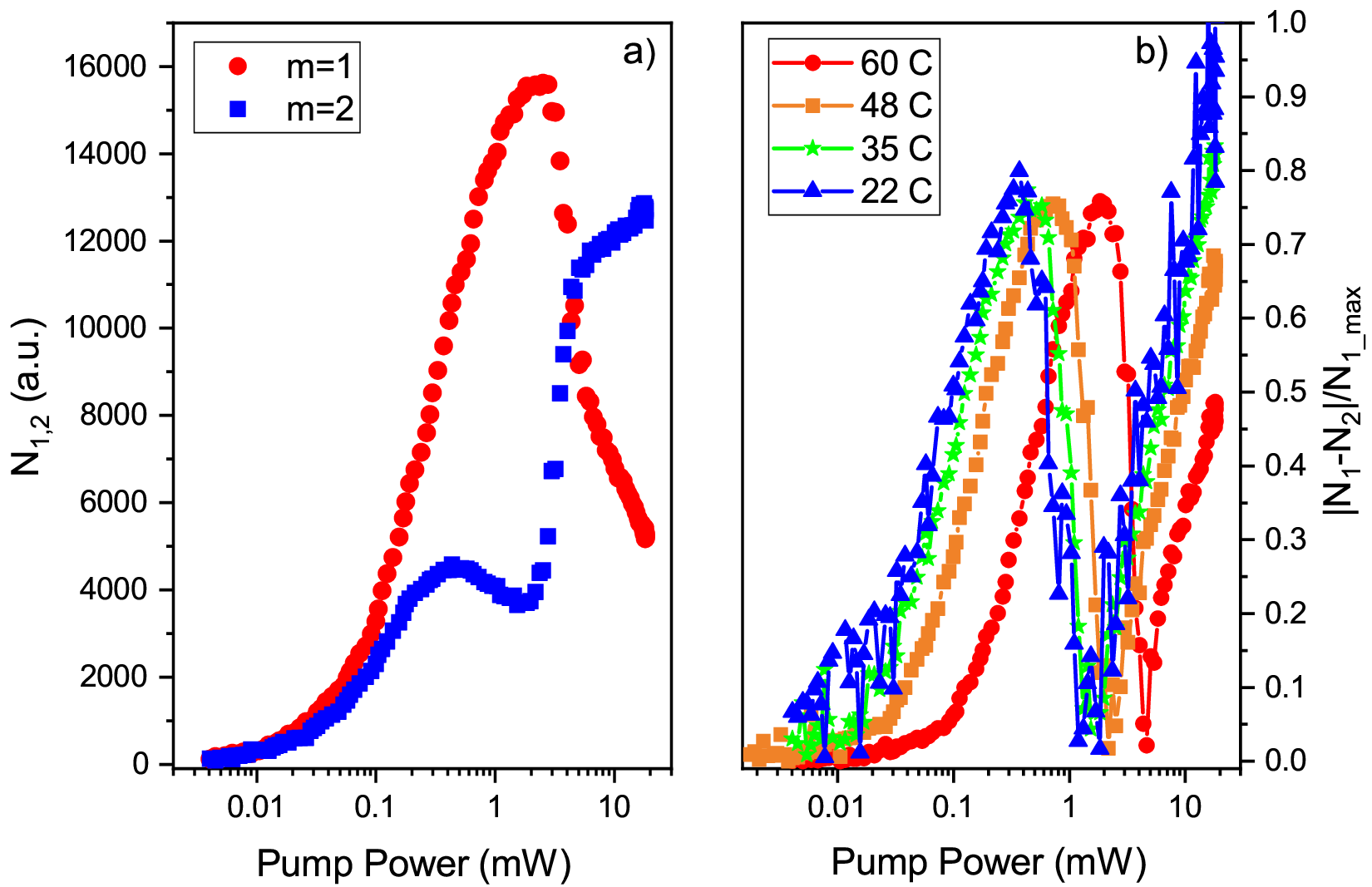}
\includegraphics[width=0.45\textwidth]{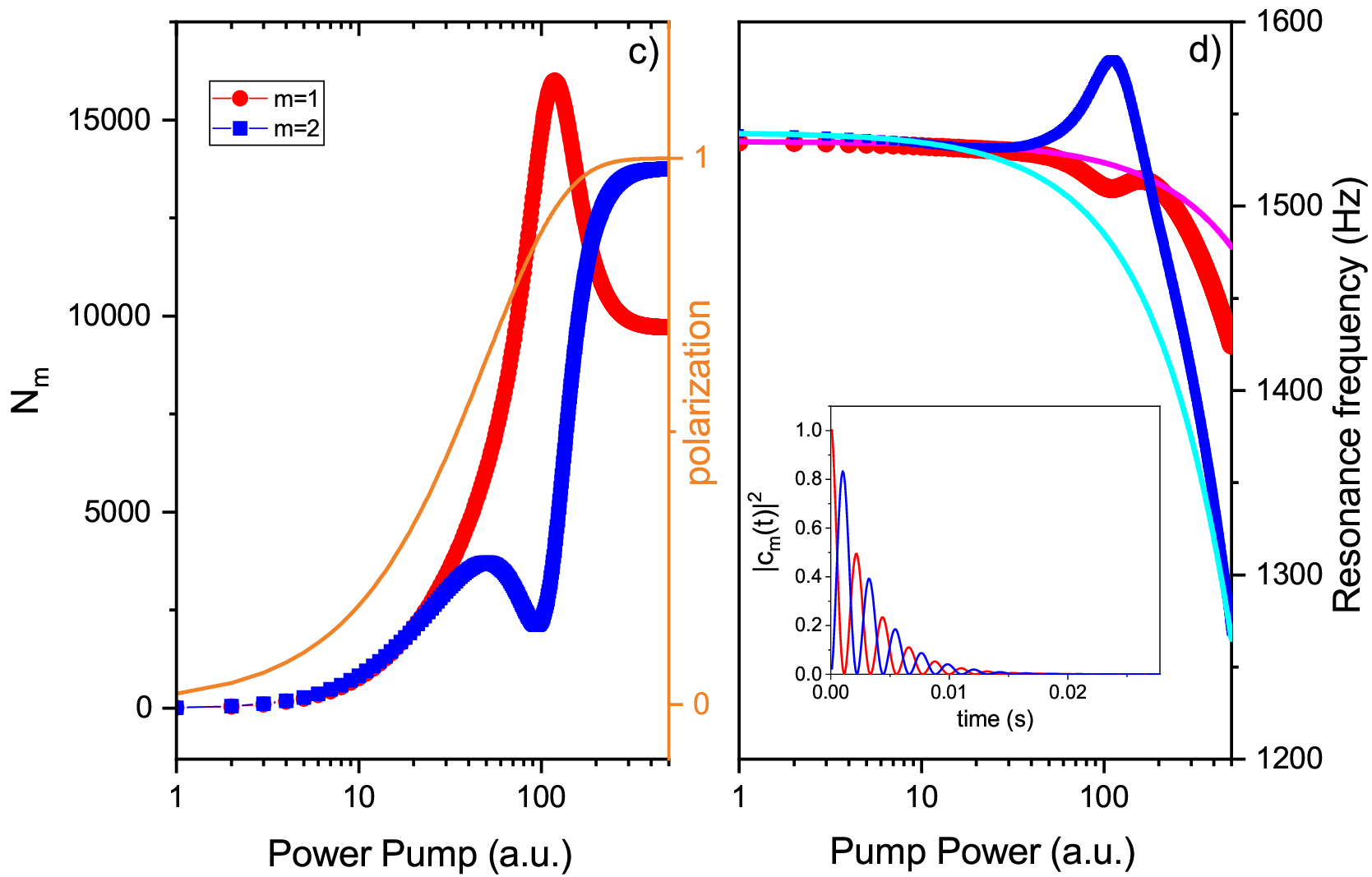}
\caption{a) Number of excitations $N_{1,2}$ for the two main spatial modes as a function of the pump power. These measurements have been taken at a temperature of $60$ $^{o}$C. b) The difference of excitations in the two modes normalized by the maximum excitation number. The signal peak shifts towards higher pump powers with increasing temperature of the atomic sample. c) Results obtained from a simple model based on Eq.\ref{eq:eq2}. The data have been calculated by using the experimental parameters with temperature of $60$ $^{o}$C, and setting $\vert J_1 \vert$ in the range $(0.5 - 0.9) \vert J_2 \vert$, and $\vert J \vert= \sqrt{\vert J_1 \vert \vert J_2 \vert}=8 \Gamma_1$. In particular, the number of excitations for the two modes as a function of the pump power is displayed together with a solid line showing the degree of alkali-metal polarization due to optical pumping. d) The resonance frequencies for the two modes as a function of pump power are displayed together with solid lines showing the experimentally derived asymptotic dependence for comparison. In addition, the inset shows the exchange of collective excitation between the two spatial modes as a function of time for a pump power value of $100$ (a.u.). }\label{fig:fig5}
\end{figure}

With our measurements, we can identify two main regimes corresponding to different values of $\vert J /\Delta \vert$: a regime where the modes are mostly independent ($\vert J / \Delta \vert <1$), and a regime where they are coupled ($\vert J / \Delta \vert >1$), which alternate when changing the pump power (for a given pump beam size). For low pump powers, when the rate of optical pumping is not greater than the rate of spin-exchange collisions, the magnetic properties across the sample are rather uniform, $\vert J / \Delta \vert <<1$, and the occupied spatial modes appear symmetric along the $z$ direction, Fig.\ref{fig:fig4}. By increasing the pump power, the inhomogeneity between the portion of the cell dominated by the pump and by the spin-exchange interaction with the noble gas, leads to values of $\vert J/\Delta  \vert >1$, see also supplemental material. In this case, we observe that the two main spin modes separate spatially as in Fig.\ref{fig:fig4}. In the presence of this potential gradient, both the resonance frequencies (similarly to an avoided crossing), and the phases of the modes shift in opposite directions. For a given pump power the relative frequency shift depends on the alkali-metal atomic density, and hence the cell temperature, as shown in the inset of Fig.\ref{fig:fig4}(b). We note that the measurements have been taken in a regime of partially polarized noble-gas, for which the total magnetization depends on the local density of polarized alkali-metal atoms, i.e. $M \propto k_{SE}N_{A}p_{A}$ (where $N_{A}$ is the density, and $p_A$ the polarization of the alkali-metal atoms).  Finally, further increase of the pump power determines a progressively more homogeneous optical pumping, and a saturation of the atomic cloud. In this regime of strong pump, the light-induced frequency shift and linewidth broadening also become significant, $\vert J /\Delta \vert <1$ and the two spatial modes evolve independently again.
To obtain further evidence of the modes' coupled dynamics, for each alkali-metal spatial mode, we have derived the total number of excitations by integrating the corresponding magnetic resonance on both the standard X and Y signals provided by the lock-in amplifier. An example of the resulting measurements is shown in Fig.\ref{fig:fig5}. For lower pump beam powers (e.g. $(P\lesssim 0.1$ mW, with a temperature of roughly $60$ $^{o}$C), excitation of the two-spatial modes has a similar growth. For higher powers, the modes show an anti-correlated non-monotonic evolution, which appears as an exchange of excitation up to roughly $7$ mW pump power. This behaviour gradually shifts towards lower pump powers for decreasing temperature of the atoms, see Fig.\ref{fig:fig5}. The dependence can be explained with the change of the on-resonance optical thickness defined for a homogeneous cloud of radius $R$ as $b_0=2N_A \sigma_{32'}R$, with $\sigma_{32'}$ being the on-resonance scattering cross-section for the pumping transition. Indeed, the mode coupling appears when the cloud is only partially optically pumped, which corresponds to lower values of the pump power for a decreased optical thickness.

We have solved a simple model based on Eqs.\ref{eq:eq2}, with in addition $\vert J \vert \propto (N_A p_A)$ and a term of gain, both depending on the variation (and saturation) of the optical pump. The agreement with the measurements is qualitatively good, as shown in Fig.\ref{fig:fig5}. The simple model also confirms that the details of the excitation signal, such as the appearance of the anti-correlated behaviour, depend on the optical saturation of the atomic cloud. The model allows us to calculate the time-dependent dynamics of the excitation \cite{comment}, and it shows a coherent exchange between the two spatial modes for the region of parameters where the anti-correlated signal is visible (inset of Fig.\ref{fig:fig5}). Note that initially, $S(\textbf{r},t\simeq 0) \simeq s_1(\textbf{r})$.
From a quantitative point of view, in these simulations we have employed a coupling coefficient $\vert J \vert \sim 8 \gamma_{1}$, where $\vert J \vert= \sqrt{\vert J_1 \vert \vert J_2 \vert}$ and $\vert J_2 \vert > \vert J_1 \vert$, which takes account of the different couplings measured for the two modes ($\omega_{Am}$ and $\gamma_{Am}$ are experimentally derived). This is roughly a factor $7$ larger than the measured values, which are however just rough estimates of the inhomogeneity due to spatial integration and the atomic thermal motion itself (see supplemental material).

Finally, we have repeated the measurements with a smaller cell containing Cs and few hundred torr of nitrogen (roughly 200 torr), in a similar temperature range. The amount of excitation for the main two modes as a function of the pump power does not appear to be anti-correlated, which we have quantified by calculating the Spearman correlation coefficient between the two modes (which is equal to $-0.37$ for the system with neon, and $+0.91$ for the system without neon at a temperature of $80$ $^{o}$C, see the supplemental material). Also, no frequency shift of the magnetic resonances is detected for the two modes, except for the pump-induced light shift as similarly observed in the neon cell measurements. At higher temperatures, when the optical thickness of the cloud equals the values typical of the neon cell, some features are detected (see supplemental materials). However, in the absence of noble gas, the gradient of the precession frequency (i.e. $\Im(J)$) is reduced, and the linewidth is increased by, at least, a factor $3$. Some frequency shifts become visible at higher values of $\Im(\Delta)$ with the same sign, and the effect of the coupling on the excitation is unnoticeable. Out of the geometric differences between the diffusive modes of these two cells, our measurements thus confirm that the neon buffer component, via spin-exchange interaction with the alkali-metal gas, plays a determinant role in enhancing the modes' coupling.

\section{Discussion}

We have experimentally studied the dynamics of multi-mode components in an atomic vapour formed by an alkali-metal and a noble gas. In these systems, the atoms do not move ballistically but can localize in stable spatial modes depending on the diffusion properties and geometry of the glass cell containing the vapour. An experimental analysis of the impact of the underlying multi-mode nature of the vapour system in a regime of parameters typical for atomic magnetometry/co-magnetometry, reveals that the occupation of a few stable spatial modes has a detectable outcome on the total spin rotation signal. The independent dynamics of each mode is crucially related to the details of the pump beam (power, dimensions, and shape), boundary, and environmental conditions, such as the presence of magnetic field \cite{Zhang2023} and light gradients. Therefore, by modifying these parameters it is possible to control the pump and probe coupling with a specific spatial mode, and to create optimized conditions of operation, as those due e.g. to a Ramsey-like effect. The latter has remarkably here demonstrated to be able to improve the stability of the lowest-order mode signal against the perturbations introduced by the pump light by almost one order of magnitude. 

Moreover, we have observed coupled dynamics between the different spatial modes. In particular, the spin precession frequency of the modes splits, as the two move spatially apart, and their phase shifts in opposite directions. This happens when a competing dynamics develops between optical pumping and spin-exchange collisions, which induces an inhomogeneity across the sample. Interestingly, this coupling leads to an exchange dynamics of collective excitations between the two modes, which we can detect as an anti-correlated dependence on the pump power, i.e. of the strength of the coupling. This coherent dynamics is especially remarkable, as it ultimately stems from the random dynamics of the diffusing spins.  

Our work demonstrates that the complexity of diffusive spin systems has excellent prospects for boosting the performances of quantum devices for sensing, and it should be especially considered in the context of precision co-magnetometry measurements \cite{asahi18}, and in schemes involving self-organizing mechanisms, e.g. the spin maser \cite{Chalupczak2015}. A systematic tailoring of the pump geometry and intensity \cite{Xiao2023}, and spatial filtering of the readout can significantly enhance the signal-to-noise and optimising the stability of the (co)-magnetometer. Metrological advantage has been shown in (nuclear) spin systems in a PT-broken phase \cite{Zhang2023}, and it is predicted near exceptional points \cite{Zhang2019}, which, as supported by our work, are also attainable in alkali-metal systems based on diffusive modes. Future developments could also include the search and study of optically controllable, localized magnetic structures \cite{Schapers2000}, also in the context of the study of phase transitions \cite{Horowicz2021}. Finally, coherent coupling of different alkali-metal spatial modes could be of interest for quantum imaging and information applications \cite{Sun19}, and interaction between the modes should be accounted for even when alkali-metal spins are used as mediators between photons and noble-gas spins.

\section{Supplemental material}

\subsection{Experiment}
The measurements described here are performed in a shielded setup, with the atomic vapour cell enclosed in a 3D-printed oven driven by an AC current modulated at a frequency of 100 KHz. The diameter of the spherical glass cell is $20$ mm. Because of the cell shape, three sets of heaters are used to heat the cell. Pumping is performed by one or a pair of circularly polarised laser beams, frequency stabilized to the $6^2$S$_{1/2}$ F=3 $\rightarrow{} 6^2$P$_{3/2}$ F'=2 transition (D2 line, $852$ nm), propagating along the direction of the bias static magnetic field, Fig.\ref{fig:fig1}.  The power of the pump beam is controlled by an acousto-optic modulator (AOM) operating in the double passage configuration. The pump beam's size is adjusted by a beam expander. The ambient magnetic field is suppressed by the use of five layers of cylindrical shields made from $2$ mm thick mu-metal with end caps (suppression factor $10^{6}$). A pair of solenoids inside the shield generates a well-controlled bias magnetic field, with a relative homogeneity at the level of $10^{-4}$ over the length of the cell. A set of Helmholtz coils orthogonal to the axis of the bias magnetic field produces the weak RF magnetic field that drives the atomic coherence. Paramagnetic Faraday rotation maps the value of the collective atomic spin onto the polarisation state of the linearly polarised probe beam \cite{Takahashi1999, Savukov2005, Chalupczak2012} that propagates orthogonally to the pump beams, and the resulting signal is measured by a lock-in amplifier referenced to the first harmonic of the RF field frequency. 

\subsection{Control via adjustment of the pump beam}
\begin{figure}
\centering
\includegraphics[width=0.45\textwidth]{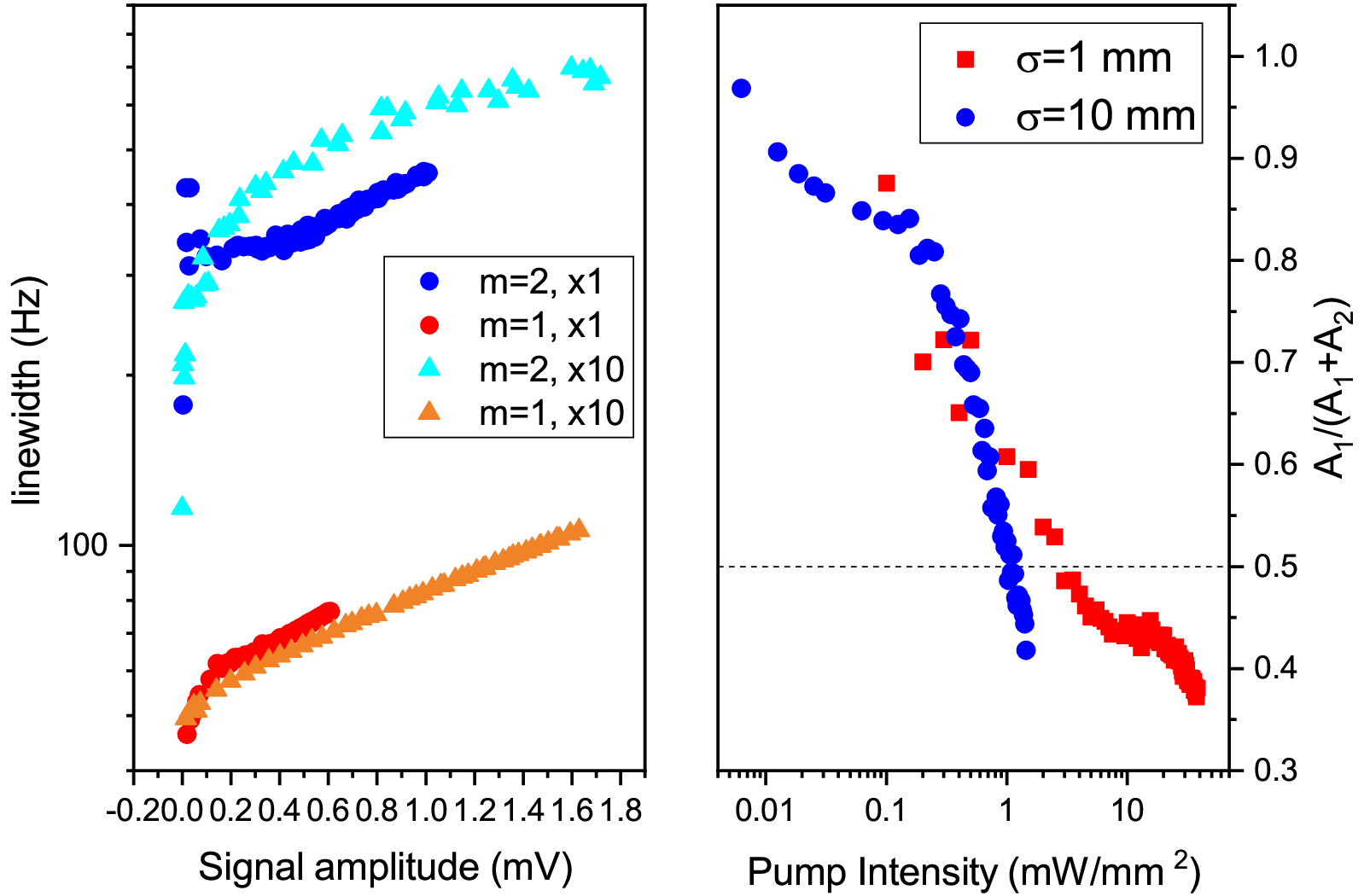}
\caption{On the left: linewidth dependence on signal amplitude for two different Gaussian beam sizes corresponding respectively to $\sigma=1$ mm (x1), and $\sigma=10$ mm (x10), measured before the atomic cell. On the right: the relative amplitude of the two main modes is affected by both intensity and size of the pump. Changing these parameters it would be possible to tune in a controlled way the contribution of each mode to the final detected signal. }\label{fig:suppl1}
\end{figure}

We have characterized the magnetic signal for two different pump beam sizes, i.e. $\sigma=1-10$ mm, where the Gaussian beam profiles are measured in front of the atomic cell. In particular, we have observed a narrowing of the linewidth of the second order mode $m=2$ by $150$ Hz as a result of the beam narrowing, for the same signal amplitudes (data above 0.3 mV in Fig.\ref{fig:suppl1}). As we have verified using non-resonant light, the distortion of the beam introduced by the spherical glass cell prevents us from obtaining a clean Gaussian profile with size $\sigma \gtrsim 6$ mm. As a result, the $m=2$ spatial mode can be either mostly or partially covered by the pump beam, while the more extended $m=1$ spatial mode is always unevenly or partially covered. Hence the different behavior of the two modes, with the $m=1$ linewidth showing no clear dependence on the pump beam size in this range of parameters. We observe also that both intensity and size of the pump beam affect the relative amplitude of the spatial modes, and by modifying one of these parameter it would be possible to change which mode is dominantly contributing to overall signal, for equal detection, see Fig.\ref{fig:suppl1}. We foresee that a more sophisticated shaping of the pump beam profile, e.g. with the use of optical axicons, and a different realization of the cell could also contribute to control the excitation of different spatial modes within the atomic cell.  

\subsection{Spatial inhomogeneity}
\begin{figure}
\centering
\includegraphics[width=0.4\textwidth]{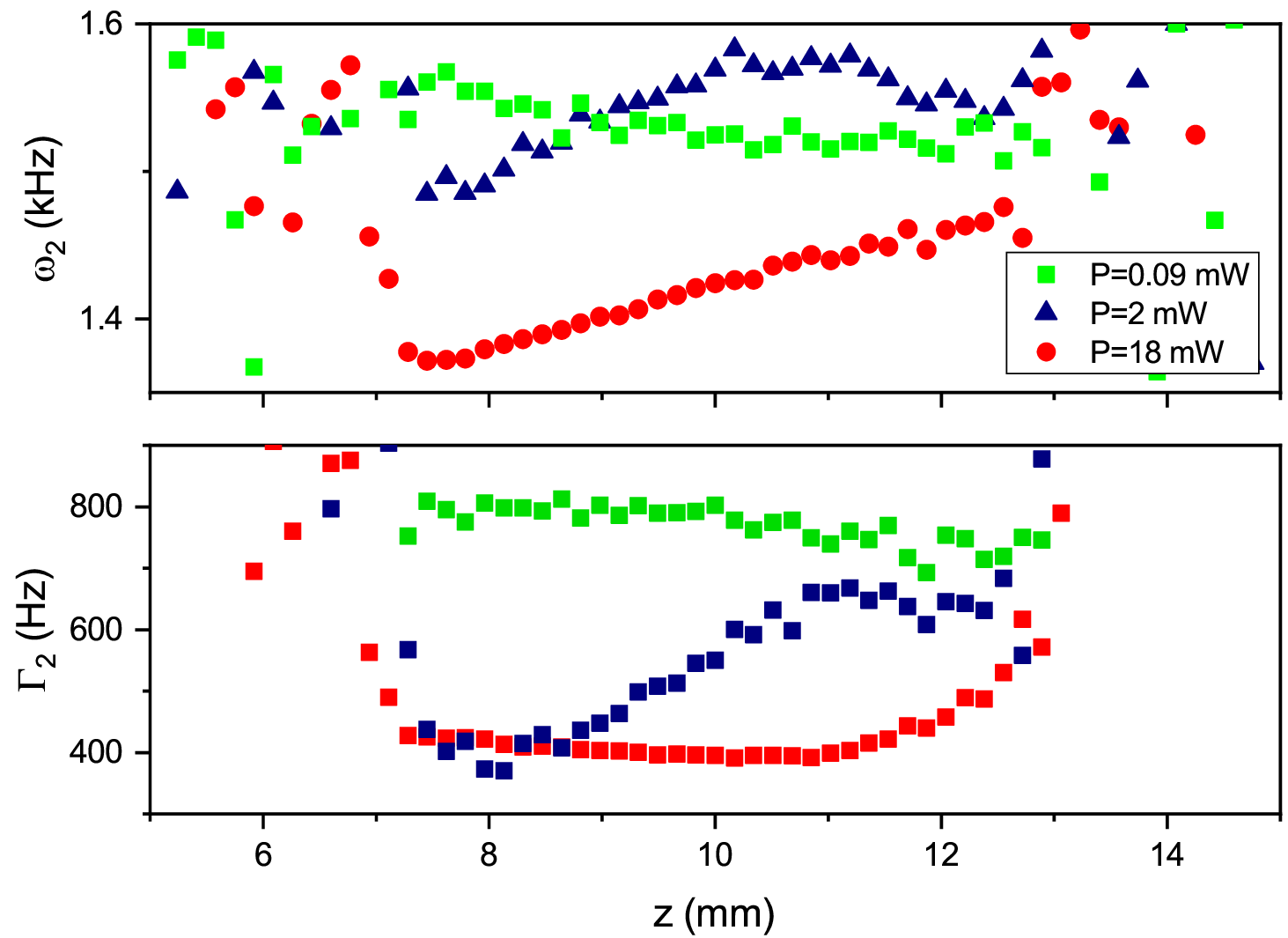}
\caption{Spin imaging showing spatial dependence of the precession frequency (above), and linewidth (below) of the magnetic resonance of mode $m=2$. This mode, being spatially localized, more evidently shows dependence on local conditions. }\label{fig:suppl2}
\end{figure}

Spatially resolved spin imaging allows us to derive an estimate of the inhomoegenous conditions through the sample ($G$). As this estimate is different for the two modes we calculate the total coupling as $\vert J \vert = \sqrt{\vert J_1 J_2 \vert}$. In the estimate of $\vert J \vert$, we account both for the inhomogeniety in the frequency (imaginary part) and in the linewidth of the two modes' magnetic signal (real part). Please note that the accuracy of this estimate is limited by the thermal motion of the atoms ($\Re(G)$ and $\Im(G)$ are correlated), and overlap with the pump beam in the transverse direction (here not seen). Example measurements are shown in Fig.\ref{fig:suppl2}.

\subsection{Atomic cell without neon buffer} 
The measurements presented in the text have been repeated in a cell containing a solid sample of cesium and a buffer of roughly $200$ torr of Nitrogen alone. The cell is built on a silicon wafer substrate and is made of two connected glass chambers with the cesium sample being hosted in one of them (INEX Microtechnology). The chamber used for the measurements has a rectangular shape with size $4\times 4$ mm$^2$ and thickness of $2$ mm. The heater is supplied by an AC current modulated at a frequency of $1$ MHz.
Analogously to the measurements in the presence of the neon buffer, the magnetic resonance peaks show composite shapes, hence we can similarly derive information on the main stationary spatial modes of the alkali-metal atoms. In order to quantify the mutual dependence and the presence of any correlation (or anti-correlation) in the excitation of the two main modes as a function of the pump power, we have calculated their Spearman correlation coefficient ($-1\leq \rho \leq 1$). For the measurements with neon at $60$ $^{o}$C, the correlation coefficient has a negative value $\rho=-0.37$ signalling the presence of a partial anti-correlation (and $\rho=-0.95$ for pump powers larger than $0.8$ mW). In the case without neon, instead, the correlation parameter has a value $\rho=+0.91$ signalling the presence of a strong positive correlation for the measurements at a temperature of $80$ $^{o}$C, and just slightly less $\rho=+0.88$ for the measurements at $96$ $^{o}$C. At a temperature of $50-60$ $^{o}$C the precession frequency does not show any shift, apart from the light-induced one. Even though these signals do not show similarities with the neon cell measurements, some features are still visible for temperatures above $80$ C$^o$, corresponding to regions where the optical thickness of the atomic cloud has reached the range of the neon cell measurements. These can be explained as a competition between optical pumping and intra-species spin exchange interactions, and will be further discussed in future publications.

\begin{figure}
\centering
\includegraphics[width=0.45\textwidth]{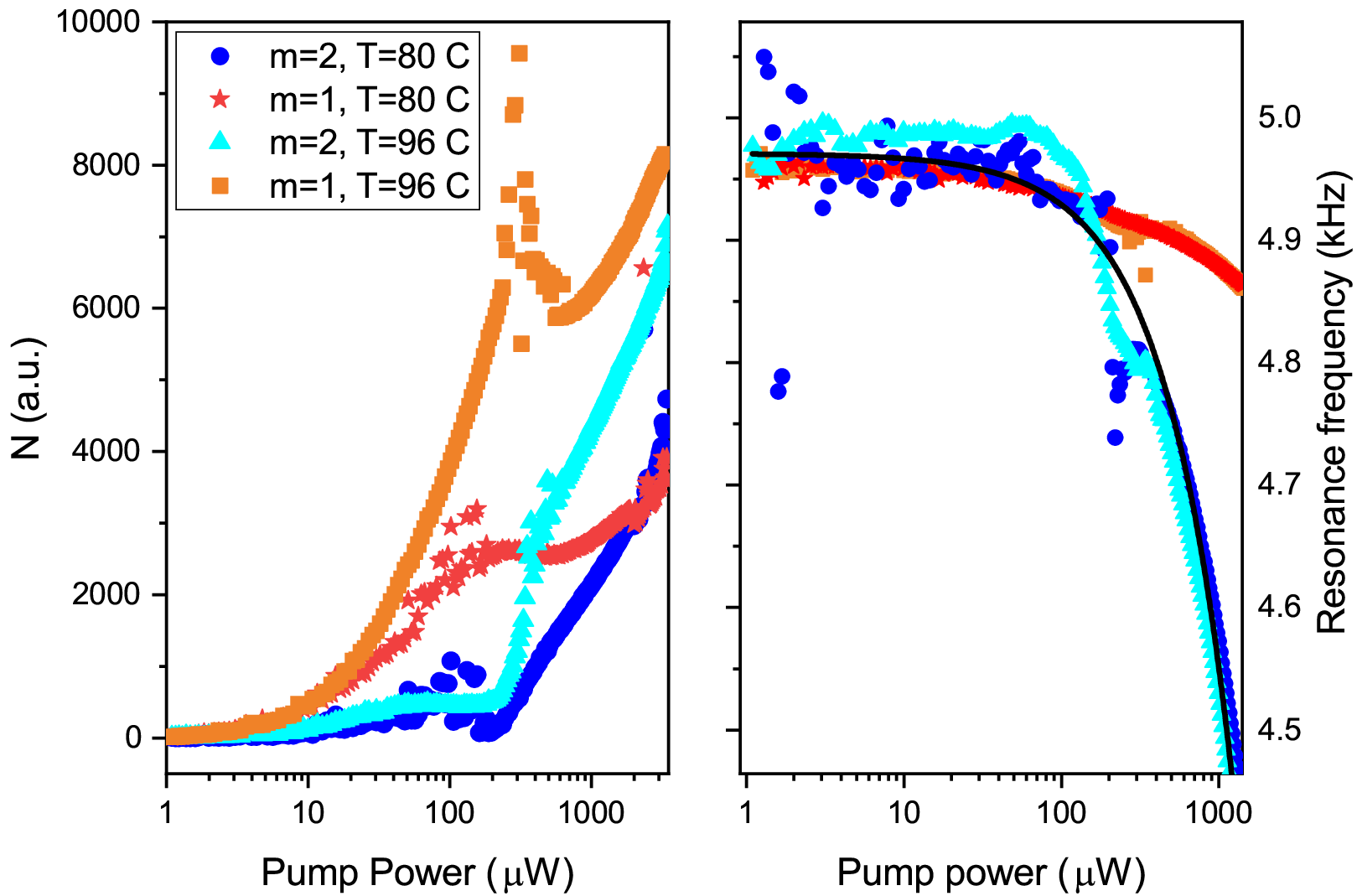}
\caption{Number of excitations (left) and resosance frequency (right) as a function of the pump power, measured for a wafer cell containing cesium and Nitrogen buffer (pressure roughly 200 torr). Black solid line is a linear fit to the data.  }\label{fig:suppl3}
\end{figure}


\begin{thebibliography}{99}
\bibitem{comment} Measurement of the dynamics during the transient would alter the magnetic properties of the modes, e.g. the linewidth, and will thus impact the estimate of the total number of excitation for each mode.
\end{thebibliography}

\section{Acknowledgements}
The work was supported by the UK Department for Business, Energy and Industrial Strategy (BEIS), and by the UK Engineering and Physical Sciences
Research Council (Grant No. EP/S000992/1). We would like to thank R. Hendricks and G. Barontini for critical reading of manuscript. We acknowledge stimulating discussions with Curt von Keyserlingk.

\section{Competing financial interests}
The authors declare no competing financial interests.

\end{document}